# Bias-controlled spectral response in GaN/AlN single-nanowire ultraviolet photodetectors

Maria Spies[1,2], Martien I. Den Hertog[1,2], Pascal Hille[3,4], Jörg Schörmann[3], Jakub Polaczyński[1,2], Bruno Gayral[1,2], Martin Eickhoff[3,4], Eva Monroy[1,5], and Jonas Lähnemann[1,5,6]

[1] University Grenoble-Alpes, 38000 Grenoble, France
[2] CNRS-Institut Néel, 25 av. des Martyrs, 38000 Grenoble, France
[3] I. Physikalisches Institut, Justus Liebig Universität Giessen, Heinrich-Buff-Ring 16, 35390 Giessen, Germany
[4] Institut für Festkörperphysik, Universität Bremen, 28359 Bremen, Germany
[5] CEA-INAC-PHELIQS, 17 av. des Martyrs, 38000 Grenoble, France
[6] Paul-Drude-Institut für Festkörperelektronik, Leibniz-Institut im Forschungsverbund Berlin e.V., Hausvogteiplatz 5-7, 10117 Berlin, Germany

**ABSTRACT:** We present a study of GaN single-nanowire ultraviolet photodetectors with an embedded GaN/AlN superlattice. The heterostructure dimensions and doping profile were designed in such a way that the application of positive or negative bias leads to an enhancement of the collection of photogenerated carriers from the GaN/AlN superlattice or from the GaN base, respectively, as confirmed by electron beam-induced current measurements. The devices display enhanced response in the ultraviolet A (≈ 330-360 nm) / B (≈ 280-330 nm) spectral windows under positive/negative bias. The result is explained by correlation of the photocurrent measurements with scanning transmission electron microscopy observations of the same single nanowire, and semi-classical simulations of the strain and band structure in one and three dimensions.

The development of semiconductor nanowires has inspired new devices where the low dimensionality and the large surface-to-volume ratio play a key role in the overall performance. Promising applications of semiconductor nanowires span from opto-,[1] electro-,[2] and spintronics[3,4] to photovoltaics,[5] energy conversion,[6–8] sensor engineering,[9,10] and biotechnology.[11] In particular, the capabilities of nanowire photodetectors have been investigated for different material systems[12–17] due to their low defect density and small electrical cross-section (i.e. low capacitance), which comes without degradation of the light absorption.[5,18,19] Compared to other materials, GaN[20–27] presents the advantage of being robust in chemically and physically extreme environments. Its direct band gap is tunable in the ultraviolet range by alloying with AlN, and the large GaN/AlN conduction band offset[28] offers interesting possibilities in terms of carrier confinement at high temperature.[29] Defect-free growth of GaN nanowires by plasma-assisted molecular-beam epitaxy (PAMBE) is an established technique,[30,31] with the possibility to form heterostructures and modulate both p- and n-type doping. The internal electric field in polar GaN/AlN nanowire heterostructures opens interesting possibilities for band profile engineering in order to modulate the spectral response, and even render it bias-dependent.

The feasibility of single-nanowire photodetectors based on GaN/AlN superlattices embedded in GaN nanowires was demonstrated by Rigutti et al.[21]. In their case, the presence of the superlattice reduced the dark current, but also decreased the responsivity by around two orders of magnitude. Their superlattice was designed to have a band-to-band transition (transition between the first confined electron and hole levels, $e_1$–$h_1$) around 430 nm ($\approx$ 2.9 eV). Temperature-dependent studies of the photocurrent, presented in ref. [21], showed that such large GaN nanodiscs (room-temperature photoluminescence measured at 2.8-3.2 eV, which corresponds to a nanodisk thickness of $\approx$ 3-5 nm) present problems for the collection of photogenerated carriers, which justifies the low responsivity. Despite of this handicap, the response from the superlattice was observed around 415 nm ($\approx$ 3 eV) in the photocurrent spectrum. Yet it was approximately three orders of magnitude weaker than the response of the GaN cap/stem. With positive bias applied to the stem, the responsivity decreased by one order of magnitude, and the response of the superlattice was masked by the background noise.

In this work, we present an improved structural design, more adapted for its use as a photodetector, which exploits the advantages associated with the internal electric field generated by the superlattice. First, we reduce the width of the GaN nanodisks to 2.5 nm to facilitate the carrier extraction. Second, we have introduced an asymmetry in the doping profile: The stem and the superlattice are undoped, whereas the cap is heavily n-type doped. This avoids the pinning of the Fermi level at the conduction band in the stem, and enlarges the regions with internal electric field in the structure. With this new architecture, we obtain an enhancement of the responsivity by more than two orders of magnitude. Furthermore, we demonstrate that the application of positive or negative bias leads to an enhancement of the collection of photogenerated carriers from the GaN/AlN superlattice (in the $\approx$ 280-330 nm ultraviolet B range) or from the GaN base (in the $\approx$ 330-360 nm ultraviolet A range), respectively. This behavior is explained using a combination of scanning transmission electron microscopy (STEM) observations, and semi-classical simulations of the strain and band structure in one and three dimensions.

## SAMPLE DESCRIPTION AND EXPERIMENTAL METHODS

The structure of the nanowires under study is depicted in Figure 1(a). The [000-1]-oriented GaN nanowires were grown on Si(111) substrates by PAMBE performed under nitrogen-rich conditions at a substrate temperature of 790°C.[32] Aluminum, gallium and germanium were supplied from standard effusion cells with beam equivalent pressures of $7\times10^{-8}$ mbar, $1.5\times10^{-7}$ mbar and $1\times10^{-10}$ mbar, respectively. The source of atomic nitrogen was operated at a power of 300 W using a nitrogen flux of 1 sccm. After the deposition of the 1600-nm-long GaN stem, a 30-period GaN/AlN superlattice is formed by periodic switching of Ga and Al fluxes. During the growth of the AlN segments, the low mobility of the impinging Al atoms gives rise to the formation of an AlN shell that envelops the GaN/AlN superlattice and the GaN stem. On the contrary, the high mobility of Ga at the high substrate temperature inhibits the formation of an additional GaN shell around the superlattice. Finally, the structure is capped with a 600-nm-long Ge-doped GaN segment ([Ge] $\approx 3\times10^{19}$ cm$^{-3}$, estimated from time-of-flight secondary ion mass spectrometry in reference samples)[33], leading to an n/n+ junction encompassing the superlattice along the nanowire axis. The structural asymmetry of the design, with a stem significantly longer than the cap, responds to an attempt to attain a homogeneous height of the GaN stems in the nanowire ensemble before the growth of the superlattice,[34] to prevent the coalescence of nanowires with displaced superlattices.[35] Figure 1(b) presents a cross-section scanning electron microscope (SEM) view of the as-grown nanowires recorded in a JEOL JSM-7001F microscope operated at 5 kV.

The periodicity of the superlattice was measured by high-resolution x-ray diffraction (HRXRD) using a PANalytical X'Pert PRO MRD system with a 4 bounce Ge(220) monochromator. The period extracted from the inter-satellite angular distance in the θ–2θ diffractogram around the (0002) reflection of the superlattice is 5.1±0.2 nm. Note that this measurement is an average of the nanowire ensemble. For a correct interpretation of the data in a single nanowire, we recorded STEM images of a contacted wire that was previously fully characterized as a photodetector. High-angle annular dark-field (HAADF) STEM was carried out using a probe-corrected FEI Titan Themis with a field-emission gun operated at 200 kV.

To electrically contact individual nanowires, the as-grown nanowires are sonicated in ethanol, and the solution is then dispersed on an array of homemade $Si_3N_4$ membranes[36] with a window size of 230 μm, compatible with STEM observations. Arrays of nitride membranes were fabricated starting from a 450-μm-thick n++ silicon (100) wafer covered on each side with a 200-nm-thick $SiO_2$ layer for additional electrical insulation and a 200-nm-thick stoichiometric $Si_3N_4$ deposited by low-pressure chemical vapor deposition. Windows and cleavage lines were opened on one side in the $Si_3N_4$ and $SiO_2$ layers by optical lithography and reactive ion etching. Subsequently, silicon and $SiO_2$ were etched through the backside in a KOH bath, until the other $Si_3N_4$ layer was reached. Then, contact pads and marker structures were defined on the front side by another optical lithography step, followed by electron beam evaporation of Ti/Au (10 nm / 50 nm) and lift-off. Single nanowires deposited on such membranes are contacted by electron beam lithography, electron beam evaporation of Ti/Al (10 nm / 120 nm) and lift-off. Figure 1(c) presents a top-view SEM image of a contacted nanowire recorded in a Zeiss Ultra+ microscope operated at 20 kV. The GaN/AlN superlattice can be identified in the image due to its darker contrast in comparison with the GaN cap and stem.

The current−voltage (I–V) characteristics were investigated both in the dark and under continuous-wave illumination with a HeCd laser (325 nm). For this purpose, the nanowires were directly connected to an Agilent 4155C semiconductor parameter analyzer and biased over a range of ±2 V. Positive bias is arbitrarily defined as indicated in Figures 1(a) and (c), i.e. higher voltage in the cap segment than in the stem. Measurements of the photocurrent as a function of optical power were made with illumination from a HeCd laser (325 nm) chopped at 73 Hz. The single nanowires were directly connected to the $10^6$ A/V transimpedance amplifier integrated in a lock-in amplifier (Stanford Research Systems SR830). To measure the spectral response of single nanowire photodetectors, we used a similar configuration, but with excitation from a 450 W xenon lamp passed through a grating monochromator.

One-dimensional and three-dimensional calculations of the band structure in the nanowires were performed using the nextnano3 software[37] with the GaN and AlN parameters described in ref. [38]. For three-dimensional calculations, the nanowire was modeled as a hexahedral prism consisting of a 150 nm long GaN section followed by a 14-period AlN/GaN stack and capped with 50 nm of GaN. The geometrical dimensions were taken from STEM measurements (radius of the GaN stem = 70 nm, AlN shell thickness = 2 nm, GaN nanodisk thickness = 2.5 nm, and AlN barrier thickness = 2.6 nm). The n-type doping density in the cap segment and the residual doping density in the GaN stem were fixed to $3\times10^{19}$ cm$^{−3}$ and $1\times10^{17}$ cm$^{−3}$, respectively. The structure was defined on a GaN substrate to provide a reference in-plane lattice parameter, and was embedded in a rectangular prism of air which allowed elastic strain relaxation. The effect of surface states was simulated by introducing a negative charge density of $2\times10^{12}$ cm$^{-2}$ at the air/nanowire side-interface, which corresponds to the density of states reported for m-plane GaN.[39] In a first stage, the three-dimensional strain distribution was calculated by minimization of the elastic energy assuming zero stress at the nanowire surface. Then, for the calculation of the band profiles, the piezoelectric fields resulting from the strain distribution were taken into account. Wavefunctions and related eigenenergies of the electron and hole states in the nanodisks were calculated by solving the Schrödinger-Poisson equations using the 8-band k.p model.

Electron beam-induced current (EBIC) imaging was carried out in a Zeiss Ultra 55 SEM operated at 20 kV with a beam current of 1.7 nA. The microscope is equipped with a Gatan SmartEBIC system. The current collected at the nanowire contacts is amplified using a SR570 current-to-voltage preamplifier, which also allows to apply an external bias voltage to the sample. Thereby, the induced current can be mapped simultaneously with the secondary electron signal from the sample.

## RESULTS AND DISCUSSION

The study presented here was validated by the observation of similar results in various nanowires with the same structure (see supplementary information). However, all the figures in this paper describe the results obtained in one of the nanowires (a typical specimen), where it was possible to perform the complete characterization as a photodetector as well as HAADF-STEM. The correlation of the photodetector performance with the microscopy study is important to confirm that it is indeed a single object, and to use its real dimensions in the band-structure simulations. Figure 1(d) shows an HAADF-STEM image of the investigated nanowire. From this micrograph, the nanowire presented here is a single nanowire resulting from the coalescence of two nanowire stems, which takes place before the growth of the superlattice. As a result, the active region is relatively large, with a diameter of ≈ 150 nm. In the active region, the GaN nanodisk and AlN barrier thicknesses are 2.5±0.3 nm and 2.6±0.3 nm, respectively, in good agreement with the data extracted from HRXRD measurements.

A semi-logarithmic plot of the I−V characteristics of the single nanowire in the dark is displayed as an inset in Figure 2. The combination of the internal electric fields induced by the heterostructure and by the n/n+ doping profile along the polar axis results in a strongly rectifying behavior. To give an idea of the homogeneity of the sample, the dark current values at −2 V and +2 V bias were in the range of 0.2-1.6 nA and 0.4-1.5 µA, respectively, for all the nanowires under study, in contrast to the up to five orders of magnitude in dispersion of these values found in ref. [35] for a less homogeneous sample. Under ultraviolet (325 nm) continuous-wave illumination, we observe an enhancement of the current by several orders of magnitude (dashed line in the inset of Figure 2 for optical power = 5 µW, spot diameter = 2 mm). Note that the ratio between photocurrent and dark current is higher under negative bias (due to the low dark current value), but the photocurrent itself is about one order of magnitude higher under positive bias conditions. Considering the nanowire surface exposed to the laser as the active photodetector area, the responsivity is as high as 150±30 kA/W at +1 V (optical power = 160 µW, spot diameter = 2 mm, the error bars account for the measurement dispersion in the various wires under study). This value is roughly 2-3 orders of magnitude larger than those in ref. [21] (0.1-2 kA/W, measured at the same irradiance, namely 5 mW/cm$^{-2}$). However, a word of caution concerning this definition of responsivities needs to be added. First, the responsivity of nanowire photodetectors is known to depend significantly on the irradiance.[24] In our case, when decreasing the laser power to 1 µW, the responsivity increases even more, up to 3.1±0.8 MA/W at +1 V and 12±3 MA/W at +2 V bias. And second, the detecting cross-section of a nanowire is known to be larger than the exposed surface,[5,18,19] so that the responsivity is overestimated. Attending to these issues, we mostly focus on photocurrent values instead of responsivities in the remainder of this manuscript and have given these values only for comparison with previous works using the same definition.

The variation of the photocurrent as a function of the impinging optical power has been studied under chopped (73 Hz) excitation at 325 nm. Note that for GaN nanowires, the chopped illumination generally leads to a reduction of the photocurrent in comparison with continuous-wave excitation.[24,35] Measurements were systematically carried out from low to high excitation power, to prevent artefacts related to persistence or temperature transients. As shown in Figure 2, the photocurrent ($I_{ph}$) increases sublinearly with the optical power ($P$) following approximately a power law $I_{ph} \sim P^{\beta}$ with $\beta < 1$ for all the measurement frequencies. This behavior was qualitatively the same for all the nanowires under study, with values of $\beta$ in the range of 0.3-0.6, and a slight difference between positive and negative bias. This sublinear response with the optical power is characteristic of the nanowire geometry, whatever the material system (GaN,[24,35,36] Si,[40] Ge,[41] ZnO,[42,43] SnO$_2$,[44] ZnTe[45]), and it is generally associated to surface-related phenomena.[20,24] Consistent with the I–V characteristics depicted in the inset of Figure 2, the photocurrent is significantly enhanced under positive bias.

There is also a photoresponse at zero bias, even though it is an order of magnitude lower than at −1 V. This signal is a confirmation of the presence of an internal electric field in the structure at zero bias. It had already been observed that the insertion of a single AlN barrier in a GaN nanowire leads to a potential asymmetry that manifests in a zero-bias photoresponse,[36,46] which can be significantly enhanced through the insertion of a GaN/AlN superlattice (photocurrent = 0.2 pA, corresponding to a responsivity of 0.1 A/W in ref. [21] under continuous-wave ultraviolet excitation). In the case of the GaN/AlN superlattice embedded in an n/n+ junction under study, the signal is further enhanced in comparison to refs. [21,35,36,46]. In our work, the photocurrent obtained under 1 mW of continuous-wave excitation with the HeCd laser (spot diameter = 2 mm) is in the 0.5-3.4 nA range, which corresponds to a responsivity in the 8-60 A/W range. Such an enhancement of the photoresponse makes it possible to observe in Figure 3 the sublinear behavior of this photovoltaic signal. In contrast to GaN photoconductors, whose photocurrent scales sublinearly with the incident optical power, the photovoltaic response in GaN p-n or Schottky photodiodes are generally linear,[47] for values of irradiance that do not introduce an important distortion of the potential profile. The linearity is associated to the different detection mechanism: in photoconductors, light absorption induces a change in the resistance/mobility of the material, which results in a modulation of the current traversing the device. In contrast, in photovoltaic devices operated at zero-bias the dark current is negligible, and photogenerated electrons and holes are separated by the internal electric field and collected at the contacts. For the nanowires under study, the maximum photon flux used in our experiments is estimated at ≈ $7 \times 10^{17}$ cm$^{-2}$s$^{-1}$, which implies a generation of carriers lower than $2 \times 10^{15}$ cm$^{-3}$ (note that this number is an estimation that takes into account the exposed nanowire exposed surface and a carrier lifetime of ≈ 50 ns, which is the photoluminescence decay time measured at 5 K), so that photogeneration should take place in a linear regime. The sublinear behavior experimentally observed in all the nanowires under study means that, once the photogenerated carriers are separated by the internal electric field, the collection process is sensitive to illumination. In other words, the device is sensitive to the photo-induced modulation of the resistance of the GaN stem/cap, which can be explained by the role of the surface states and the large surface-to-volume ratio.

The above-described study of the variation of the photocurrent as a function of the incident optical power is a prerequisite for a correct interpretation of spectral response measurements. The recorded photocurrent spectra are modulated by the spectral intensity of the lamp, and the correction must take into account the sublinearity of the devices. Corrected photocurrent spectra for various bias values are presented in Figure 3. The spectra are normalized and vertically shifted for clarity (the response under positive bias is always at least one order of magnitude higher than under negative bias). Under negative bias, the spectral response is maximum in the range of 330-360 nm. In contrast, under positive bias, the maximum response is obtained in the 310-330 nm band. Note that the spectral response is presented here in a linear scale, since we are interested in the behavior in the ultraviolet region. Above 400 nm, the response of all the devices is below the detection limit of our measurement system, whatever the bias.

This bias modulated change in the contributions to the photocurrent agrees with the trend in the power law coefficient β observed in Figure 2. The sub-linear behavior of the photocurrent with optical power is most pronounced for negative bias ($\beta = 0.32$), intermediate for zero bias ($\beta = 0.42$) and significantly reduced for positive bias ($\beta = 0.58$). In other words, when the superlattice contributes to the photocurrent, the influence of the surface states on the conductivity is not as pronounced.

The different behavior of the structure under positive and negative bias can be understood by analysis of the potential profile along the nanowire growth axis. Figure 4 shows the band diagram in the nanowire calculated using the Nextnano3 software in one dimension, for a nanodisk/barrier thickness of 2.5 nm / 2.6 nm, considering a residual doping of $5\times10^{17}$ cm$^{-3}$ and a cap layer doped at [Ge] = $3\times10^{19}$ cm$^{-3}$. The figure presents the band diagram under negative bias (–$V_B$, top image), at zero bias (in the center), and under positive bias (+$V_B$, bottom image). At zero bias, the band profile presents a triangular shape induced by the negative charge sheet at the interface between the GaN stem and the superlattice. This fixed charge is due to the difference in polarization between GaN and the superlattice. It creates a large depletion region (>200 nm) in the GaN stem and an internal electric field along the whole superlattice, which is almost fully depleted. At the interface between the superlattice and the GaN cap, on the contrary, the difference in polarization generates a positively charged sheet, which pins the Fermi level at the conduction band. The fact that the cap is heavily doped leads to flat bands all along the cap section. In summary, the sample contains two areas with internal electric field: the superlattice, with electric field pointing towards [0001], and the depletion region in the stem, with electric field pointing towards [000–1]. The application of bias determines the dominant electric field. Under negative bias (Figure 4(a)), the bands flatten along the superlattice, and predominantly electrons absorbed in the GaN stem are collected.

Under positive bias (Figure 4(c)), the depletion region in the stem shrinks and collection of carriers absorbed in the superlattice is favored. At the same time, the barrier for carriers injected from the contacts is higher for negative bias (Figure 4(a)) than for positive bias (Figure 4(c)), which explains the asymmetry of the dark currents in the inset to Figure 2.

The response from different regions of the nanowires under reverse and forward bias should be directly visible in maps of the EBIC signal. Such maps recorded on the investigated nanowire are presented in Figures 4(d) and 4(e), respectively. In EBIC, electron-hole pairs separated by the internal or applied electric fields are collected through the contacts. Scanning the electron beam as excitation source over the sample results in a map of the origin of the photocurrent, i.e. of the depletion regions in the sample. As expected from the spectral response measurements together with the band structure simulations, the signal for positive bias in Figure 4(e) largely coincides with the location of the superlattice. Under negative bias (Figure 4(d)), the signal originates from the stem of the nanowire, below and close to the metal contact. The location and shape of the EBIC signal in this case, points to the stem contact being rather a Schottky contact, which is reasonable since the stem is non-intentionally doped. Some additional weak signal components correlate with morphological features in the SEM image, notably with the edge of the metal contacts.

The band diagram at zero bias presented in Figure 4(b) is rather symmetric with respect to the heterointerface between the GaN stem and the superlattice. Therefore, it is in contradiction with the significant photoresponse observed at zero bias. To understand this experimental evidence, calculations must consider the three-dimensional nature of the object. We have performed three-dimensional calculations of the band diagram using the nextnano3 software (see details in the methods), with the results presented in Figure 5(a-c), where (a) is the structure used for the calculation, and (b) is a cross-section of the band diagram in the center of the wire viewed along <1–100>, and (c) is the band diagram along the central axis of the wire. In spite of the large diameter of the wire, the presence of surface states leads to a complete depletion of the nanowire stem. As a result, the band profile is asymmetric at zero bias. Comparing with Figure 4, we could say that surface states introduce a "bias offset", i.e. at zero bias the bands resemble the positive bias configuration in planar structures (Figure 4(c)).

To understand the photoresponse of the superlattice in the 280-330 nm band, we have performed one-dimensional and three-dimensional calculations of the electronic levels in the GaN nanodisks, using the nextnano3 8-band k.p solver. Three dimensional calculations are important to visualize the in-plane separation of the electron and the hole due to the shear component of the strain.[48] Figure 5(e) represents the location of the wavefunctions of the first electron ($e_1$) and hole ($h_1$) levels in a nanodisk located in the center of the GaN/AlN stack (cross-section view along <1–100>). Even if the wavefunctions would present a certain overlap along <0001>, their in-plane separation renders the transition extremely improbable. Therefore, the absorption should be associated to the $e_2$–$h_2$ transition, whose observation is also favored by the higher density of states, and the higher collection probability for the photoexcited electron to contribute to the photocurrent.

A precise calculation of the $e_2$–$h_2$ transition energy in three dimensions is difficult due to the high amount of laterally confined levels in GaN nanodisks with such a large diameter, which leads to enormous computation time. However, as the second level is higher in energy, it also experiences a reduced radial confinement, and one-dimensional calculations should provide a reasonable approach. Figure 5(d) presents the confined levels resulting from a one-dimensional simulation, which leads to $e_1$–$h_1$ = 2.94 eV (=421 nm) and $e_2$–$h_2$ = 3.90 eV (=318 nm), consistent with our assumption that $e_2$–$h_2$ is responsible for the photocurrent enhancement in the 280-330 nm spectral region.

## CONCLUSIONS

In summary, we have presented a study of GaN single-nanowire photodetectors with an embedded superlattice consisting of 30 periods of GaN/AlN. The stem and the superlattice are undoped whereas the cap is heavily doped n-type. The heterostructure dimensions and doping profile were designed to exploit the advantages associated with the internal electric field generated by the superlattice. With this architecture, we obtain a significant enhancement of both the response under positive bias and the photovoltaic (zero-bias) response with respect to previous results. Furthermore, we demonstrate that the application of positive or negative bias leads to the dominant collection of photogenerated carriers from the GaN/AlN superlattice (in the ≈ 280-330 nm range) or from the GaN stem (in the ≈ 330-360 nm range), respectively. This bias-dependent collection mechanism is consistent with EBIC measurements, and can be explained form the correlation of the photocurrent measurements with STEM observations of the same single nanowire and semi-classical simulations of the strain and band structure in one and three dimensions.

## ACKNOWLEDGEMENTS

This work is supported by the EU ERC-StG "TeraGaN" (#278428) and ANR-COSMOS

(ANR-12-JS10-0002) projects. We benefited from access to the Nanocharacterization platform (PFNC) in CEA Minatec Grenoble. Membrane production and nanowire contacting were carried out at the Nanofab cleanroom of Institut Néel. Thanks are due to Bruno Fernandez and Jean-François Motte for their technical support.

**REFERENCES**

(1) Huang, M. H.; Mao, S.; Feick, H.; Yan, H.; Wu, Y.; Kind, H.; Weber, E.; Russo, R.; Yang, P. Room-Temperature Ultraviolet Nanowire Nanolasers. *Science* **2001**, *292*, 1897–1899.
(2) Lu, W.; Xie, P.; Lieber, C. M. Nanowire Transistor Performance Limits and Applications. *IEEE Trans. Electron Devices* **2008**, *55*, 2859–2876.
(3) van Weert, M. H. M.; Akopian, N.; Perinetti, U.; van Kouwen, M. P.; Algra, R. E.; Verheijen, M. A.; Bakkers, E. P. A. M.; Kouwenhoven, L. P.; Zwiller, V. Selective Excitation and Detection of Spin States in a Single Nanowire Quantum Dot. *Nano Lett.* **2009**, *9*, 1989–1993.
(4) Hegde, M.; Farvid, S. S.; Hosein, I. D.; Radovanovic, P. V. Tuning Manganese Dopant Spin Interactions in Single GaN Nanowires at Room Temperature. *ACS Nano* **2011**, *5*, 6365–6373.
(5) Krogstrup, P.; Jørgensen, H. I.; Heiss, M.; Demichel, O.; Holm, J. V.; Aagesen, M.; Nygard, J.; Fontcuberta i Morral, A. Single-Nanowire Solar Cells beyond the Shockley–Queisser Limit. *Nat. Photonics* **2013**, *7*, 306–310.
(6) Hochbaum, A. I.; Yang, P. Semiconductor Nanowires for Energy Conversion. *Chem. Rev.* **2010**, *110*, 527–546.
(7) Kibria, M. G.; Nguyen, H. P. T.; Cui, K.; Zhao, S.; Liu, D.; Guo, H.; Trudeau, M. L.; Paradis, S.; Hakima, A.-R.; Mi, Z. One-Step Overall Water Splitting under Visible Light Using Multiband InGaN/GaN Nanowire Heterostructures. *ACS Nano* **2013**, *7*, 7886–7893.
(8) Jung, H. S.; Hong, Y. J.; Li, Y.; Cho, J.; Kim, Y.-J.; Yi, G.-C. Photocatalysis Using GaN Nanowires. *ACS Nano* **2008**, *2*, 637–642.
(9) Cui, Y. Nanowire Nanosensors for Highly Sensitive and Selective Detection of Biological and Chemical Species. *Science* **2001**, *293*, 1289–1292.
(10) Wallys, J.; Teubert, J.; Furtmayr, F.; Hofmann, D. M.; Eickhoff, M. Bias-Enhanced Optical pH Response of Group III–Nitride Nanowires. *Nano Lett.* **2012**, *12*, 6180–6186.
(11) Li, J.; Han, Q.; Zhang, Y.; Zhang, W.; Dong, M.; Besenbacher, F.; Yang, R.; Wang, C. Optical Regulation of Protein Adsorption and Cell Adhesion by Photoresponsive GaN Nanowires. *ACS Appl. Mater. Interfaces* **2013**, *5*, 9816–9822.
(12) Soci, C.; Zhang, A.; Xiang, B.; Dayeh, S. A.; Aplin, D. P. R.; Park, J.; Bao, X. Y.; Lo, Y. H.; Wang, D. ZnO Nanowire UV Photodetectors with High Internal Gain. *Nano Lett.* **2007**, *7*, 1003–1009.
(13) Wang, J.; Gudiksen, M. S.; Duan, X.; Cui, Y.; Lieber, C. M. Highly Polarized Photoluminescence and Photodetection from Single Indium Phosphide Nanowires. *Science* **2001**, *293*, 1455–1457.
(14) Bae, J.; Kim, H.; Zhang, X.-M.; Dang, C. H.; Zhang, Y.; Choi, Y. J.; Nurmikko, A.; Wang, Z. L. Si Nanowire Metal–insulator–semiconductor Photodetectors as Efficient Light Harvesters. *Nanotechnology* **2010**, *21*, 095502.
(15) Dai, X.; Zhang, S.; Wang, Z.; Adamo, G.; Liu, H.; Huang, Y.; Couteau, C.; Soci, C. GaAs/AlGaAs Nanowire Photodetector. *Nano Lett.* **2014**, *14*, 2688–2693.
(16) Cao, L.; Park, J.-S.; Fan, P.; Clemens, B.; Brongersma, M. L. Resonant Germanium Nanoantenna Photodetectors. *Nano Lett.* **2010**, *10*, 1229–1233.
(17) Seyedi, M. A.; Yao, M.; O'Brien, J.; Wang, S. Y.; Dapkus, P. D. Large Area, Low

Capacitance, GaAs Nanowire Photodetector with a Transparent Schottky Collecting Junction. *Appl. Phys. Lett.* **2013**, *103*, 251109.

(18) Diedenhofen, S. L.; Janssen, O. T. A.; Grzela, G.; Bakkers, E. P. A. M.; Gómez Rivas, J. Strong Geometrical Dependence of the Absorption of Light in Arrays of Semiconductor Nanowires. *ACS Nano* **2011**, *5*, 2316–2323.

(19) Xu, Y.; Gong, T.; Munday, J. N. The Generalized Shockley-Queisser Limit for Nanostructured Solar Cells. *Sci. Rep.* **2015**, *5*, 13536.

(20) Calarco, R.; Marso, M.; Richter, T.; Aykanat, A. I.; Meijers, R.; v.d. Hart, A.; Stoica, T.; Lüth, H. Size-Dependent Photoconductivity in MBE-Grown GaN−Nanowires. *Nano Lett.* **2005**, *5*, 981–984.

(21) Rigutti, L.; Tchernycheva, M.; De Luna Bugallo, A.; Jacopin, G.; Julien, F. H.; Zagonel, L. F.; March, K.; Stephan, O.; Kociak, M.; Songmuang, R. Ultraviolet Photodetector Based on GaN/AlN Quantum Disks in a Single Nanowire. *Nano Lett.* **2010**, *10*, 2939–2943.

(22) Sanford, N. A.; Blanchard, P. T.; Bertness, K. A.; Mansfield, L.; Schlager, J. B.; Sanders, A. W.; Roshko, A.; Burton, B. B.; George, S. M. Steady-State and Transient Photoconductivity in c-Axis GaN Nanowires Grown by Nitrogen-Plasma-Assisted Molecular Beam Epitaxy. *J. Appl. Phys.* **2010**, *107*, 034318.

(23) González-Posada, F.; Songmuang, R.; Den Hertog, M.; Monroy, E. Environmental Sensitivity of N-I-N and Undoped Single GaN Nanowire Photodetectors. *Appl. Phys. Lett.* **2013**, *102*, 213113.

(24) González-Posada, F.; Songmuang, R.; Den Hertog, M.; Monroy, E. Room-Temperature Photodetection Dynamics of Single GaN Nanowires. *Nano Lett.* **2012**, *12*, 172–176.

(25) Sanford, N. A.; Robins, L. H.; Blanchard, P. T.; Soria, K.; Klein, B.; Eller, B. S.; Bertness, K. A.; Schlager, J. B.; Sanders, A. W. Studies of Photoconductivity and Field Effect Transistor Behavior in Examining Drift Mobility, Surface Depletion, and Transient Effects in Si-Doped GaN Nanowires in Vacuum and Air. *J. Appl. Phys.* **2013**, *113*, 174306.

(26) Wang, X.; Zhang, Y.; Chen, X.; He, M.; Liu, C.; Yin, Y.; Zou, X.; Li, S. Ultrafast, Superhigh Gain Visible-Blind UV Detector and Optical Logic Gates Based on Nonpolar a-Axial GaN Nanowire. *Nanoscale* **2014**, *6*, 12009–12017.

(27) Chen, R. S.; Tsai, H. Y.; Huang, Y. S.; Chen, Y. T.; Chen, L. C.; Chen, K. H. Photoconduction Efficiencies and Dynamics in GaN Nanowires Grown by Chemical Vapor Deposition and Molecular Beam Epitaxy: A Comparison Study. *Appl. Phys. Lett.* **2012**, *101*, 113109.

(28) Tchernycheva, M.; Nevou, L.; Doyennette, L.; Julien, F.; Warde, E.; Guillot, F.; Monroy, E.; Bellet-Amalric, E.; Remmele, T.; Albrecht, M. Systematic Experimental and Theoretical Investigation of Intersubband Absorption in GaN/AlN Quantum Wells. *Phys. Rev. B* **2006**, *73*, 125347.

(29) Holmes, M. J.; Choi, K.; Kako, S.; Arita, M.; Arakawa, Y. Room-Temperature Triggered Single Photon Emission from a III-Nitride Site-Controlled Nanowire Quantum Dot. *Nano Lett.* **2014**, *14*, 982–986.

(30) Yoshizawa, M.; Kikuchi, A.; Mori, M.; Fujita, N.; Kishino, K. Growth of Self-Organized GaN Nanostructures on Al2O3(0001) by RF-Radical Source Molecular Beam Epitaxy. *Jpn. J. Appl. Phys.* **1997**, *36*, L459.

(31) Sanchez-Garcia, M. A.; Calleja, E.; Monroy, E.; Sanchez, F. J.; Calle, F.; Muñoz, E.; Beresford, R. The Effect of the III/V Ratio and Substrate Temperature on the Morphology and Properties of GaN- and AlN-Layers Grown by Molecular Beam Epitaxy on Si(1 1 1). *J. Cryst. Growth* **1998**, *183*, 23–30.

(32) Furtmayr, F.; Teubert, J.; Becker, P.; Conesa-Boj, S.; Morante, J. R.; Chernikov, A.; Schäfer, S.; Chatterjee, S.; Arbiol, J.; Eickhoff, M. Carrier Confinement in GaN/AlxGa1−xN Nanowire Heterostructures (0<x≤1). *Phys. Rev. B* **2011**, *84*, 205303.

(33) Schörmann, J.; Hille, P.; Schäfer, M.; Müßener, J.; Becker, P.; Klar, P. J.; Kleine-Boymann, M.; Rohnke, M.; de la Mata, M.; Arbiol, J.; *et al.* Germanium Doping of Self-Assembled GaN Nanowires Grown by Plasma-Assisted Molecular Beam Epitaxy. *J.*

(34) Fernández-Garrido, S.; Zettler, J. K.; Geelhaar, L.; Brandt, O. Monitoring the Formation of Nanowires by Line-of-Sight Quadrupole Mass Spectrometry: A Comprehensive Description of the Temporal Evolution of GaN Nanowire Ensembles. *Nano Lett.* **2015**, *15*, 1930–1937.

(35) Lähnemann, J.; Den Hertog, M.; Hille, P.; de la Mata, M.; Fournier, T.; Schörmann, J.; Arbiol, J.; Eickhoff, M.; Monroy, E. UV Photosensing Characteristics of Nanowire-Based GaN/AlN Superlattices. *Nano Lett.* **2016**, *16*, 3260–3267.

(36) den Hertog, M. I.; González-Posada, F.; Songmuang, R.; Rouviere, J. L.; Fournier, T.; Fernandez, B.; Monroy, E. Correlation of Polarity and Crystal Structure with Optoelectronic and Transport Properties of GaN/AlN/GaN Nanowire Sensors. *Nano Lett.* **2012**, *12*, 5691–5696.

(37) Birner, S.; Zibold, T.; Andlauer, T.; Kubis, T.; Sabathil, M.; Trellakis, A.; Vogl, P. Nextnano: General Purpose 3-D Simulations. *IEEE Trans. Electron Devices* **2007**, *54*, 2137–2142.

(38) Kandaswamy, P. K.; Guillot, F.; Bellet-Amalric, E.; Monroy, E.; Nevou, L.; Tchernycheva, M.; Michon, A.; Julien, F. H.; Baumann, E.; Giorgetta, F. R.; et al. GaN/AlN Short-Period Superlattices for Intersubband Optoelectronics: A Systematic Study of Their Epitaxial Growth, Design, and Performance. *J. Appl. Phys.* **2008**, *104*, 093501.

(39) Bertelli, M.; Löptien, P.; Wenderoth, M.; Rizzi, A.; Ulbrich, R.; Righi, M.; Ferretti, A.; Martin-Samos, L.; Bertoni, C.; Catellani, A. Atomic and Electronic Structure of the Nonpolar GaN(1-100) Surface. *Phys. Rev. B* **2009**, *80*, 115324.

(40) Zhang, A.; You, S.; Soci, C.; Liu, Y.; Wang, D.; Lo, Y.-H. Silicon Nanowire Detectors Showing Phototransistive Gain. *Appl. Phys. Lett.* **2008**, *93*, 121110.

(41) Kim, C.-J.; Lee, H.-S.; Cho, Y.-J.; Kang, K.; Jo, M.-H. Diameter-Dependent Internal Gain in Ohmic Ge Nanowire Photodetectors. *Nano Lett.* **2010**, *10*, 2043–2048.

(42) Soci, C.; Zhang, A.; Xiang, B.; Dayeh, S. A.; Aplin, D. P. R.; Park, J.; Bao, X. Y.; Lo, Y. H.; Wang, D. ZnO Nanowire UV Photodetectors with High Internal Gain. *Nano Lett.* **2007**, *7*, 1003–1009.

(43) Chen, M.-W.; Chen, C.-Y.; Lien, D.-H.; Ding, Y.; He, J.-H. Photoconductive Enhancement of Single ZnO Nanowire through Localized Schottky Effects. *Opt. Express* **2010**, *18*, 14836.

(44) Hu, L.; Yan, J.; Liao, M.; Wu, L.; Fang, X. Ultrahigh External Quantum Efficiency from Thin $SnO_2$ Nanowire Ultraviolet Photodetectors. *Small* **2011**, *7*, 1012–1017.

(45) Cao, Y. L.; Liu, Z. T.; Chen, L. M.; Tang, Y. B.; Luo, L. B.; Jie, J. S.; Zhang, W. J.; Lee, S. T.; Lee, C. S. Single-Crystalline ZnTe Nanowires for Application as High-Performance Green/Ultraviolet Photodetector. *Opt. Express* **2011**, *19*, 6100.

(46) Hertog, M. den; Songmuang, R.; Gonzalez-Posada, F.; Monroy, E. Single GaN-Based Nanowires for Photodetection and Sensing Applications. *Jpn. J. Appl. Phys.* **2013**, *52*, 11NG01.

(47) Monroy, E.; Omn s, F.; Calle, F. Wide-Bandgap Semiconductor Ultraviolet Photodetectors. *Semicond. Sci. Technol.* **2003**, *18*, R33–R51.

(48) Beeler, M.; Lim, C. B.; Hille, P.; Bleuse, J.; Schörmann, J.; de la Mata, M.; Arbiol, J.; Eickhoff, M.; Monroy, E. Long-Lived Excitons in GaN/AlN Nanowire Heterostructures. *Phys. Rev. B* **2015**, *91*, 205440.

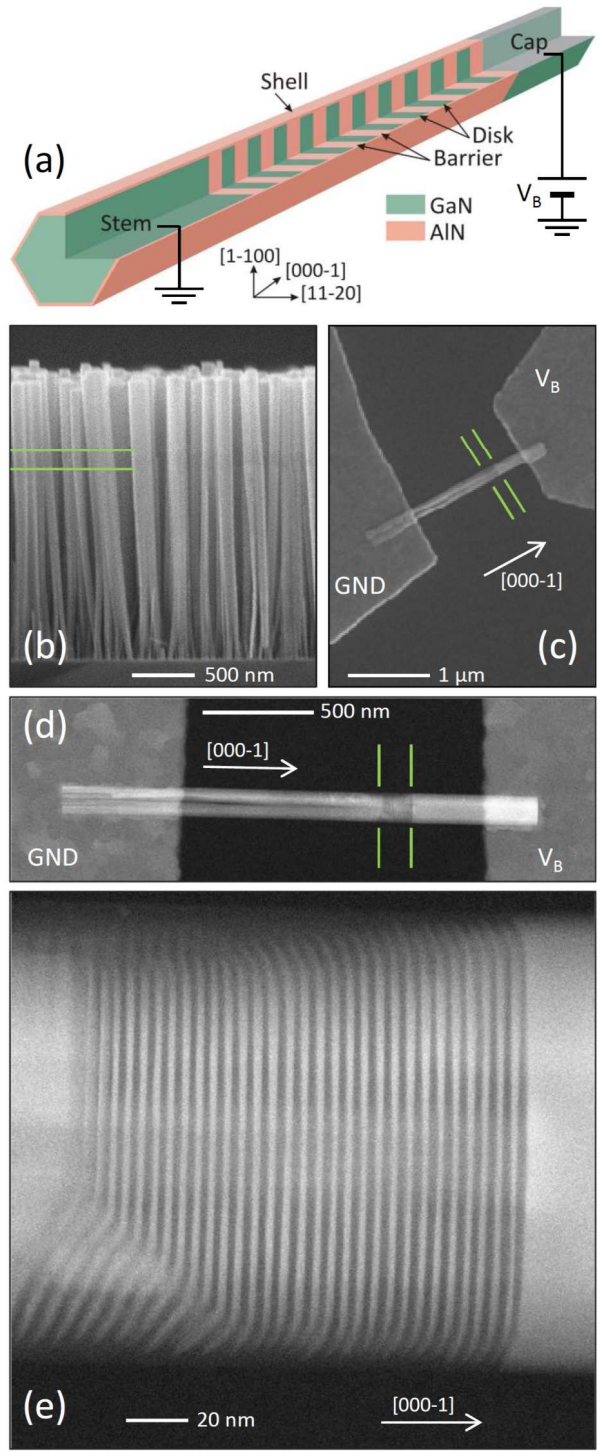

**Figure 1.** (a) Schematic description of the nanowire heterostructure. The voltage convention is indicated, with bias applied to the GaN cap ($V_B$) whereas the GaN stem is grounded (GND). (b) Cross-section SEM image of the as-grown nanowire ensemble. The GaN/AlN superlattice, outlined with green horizontal lines, appears with a darker contrast. (c) SEM image of a contacted nanowire. The GaN/AlN superlattice, between the green parallel lines, appears with a darker contrast. The voltage convention is indicated, with bias applied to the GaN cap ($V_B$) whereas the GaN stem is grounded (GND). (d) HAADF-STEM image of the same nanowire. (e) Zoomed HAADF-STEM image revealing the superlattice structure. Clear areas correspond to GaN, darker contrast to AlN.

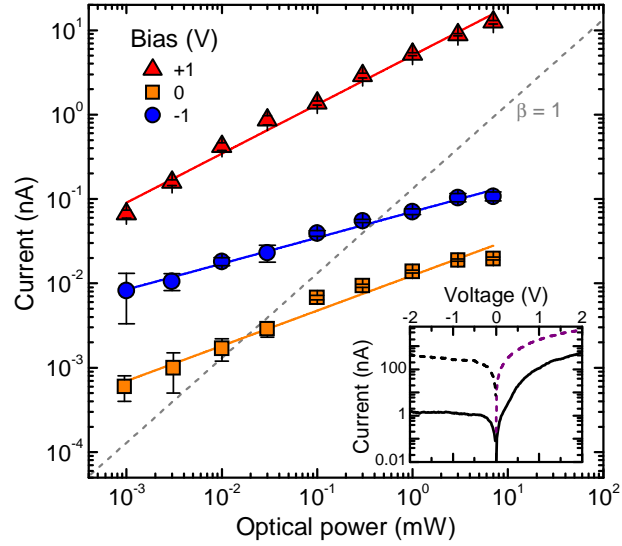

**Figure 2.** Variation of the photocurrent (*I*) as a function of the impinging optical power (*P*), measured with a HeCd laser (325 nm). Results for $V_B = +1$ V, 0 V and $-1$ V are presented. The error bars in the diagram represent the noise in the measurements. The photocurrent follows an $I \sim P^\beta$ power law (solid lines) with $\beta = 0.58 \pm 0.02$ under positive bias and $\beta = 0.31 \pm 0.02$ under negative bias. At zero bias, the response is still sublinear, with $\beta = 0.42 \pm 0.03$. The gray dashed line represents the slope for $\beta = 1$, i.e. linear behavior. Inset: Current-voltage characteristic of the nanowire under study, recorded in the dark (solid line) and under illumination (dashed line) with a continuous-wave HeCd laser (325 nm, power = 5 µW, spot diameter 2 mm).

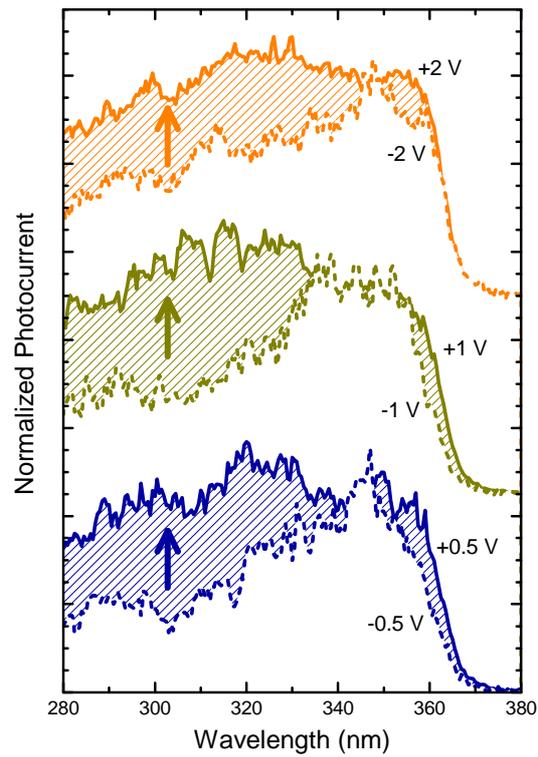

**Figure 3.** Spectral response of a single nanowire at various values of bias voltages (±0.5 V, ±1 V, ±2 V). Data are corrected by the Xe-lamp emission spectrum taking the sublinear power dependence of the NW into account. Measurements are normalized and vertically shifted for clarity. Solid (dashed) lines correspond to positive (negative) bias. Shadowed areas outline the difference in the response between positive and negative bias.

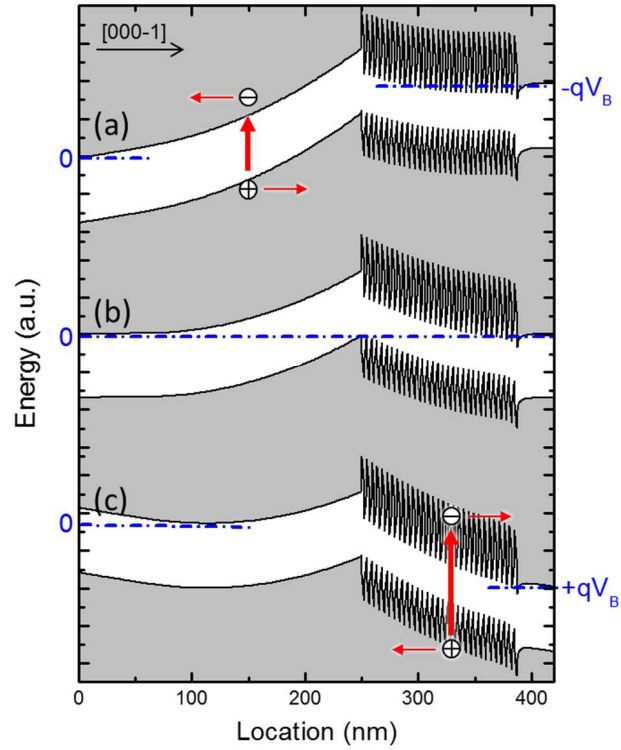

**Figure 4.** Description of the evolution of the band diagram under bias: (a) negative bias, (b) zero bias, and (c) positive bias. The data is the result of one-dimensional calculations of the nanowire heterostructure, in presence of an external electric field. The Fermi level(s) is (are) indicated by a dash-dotted blue line in each diagram. EBIC heat maps at (d) –2 V bias and (e) +2 V bias superimposed with an SEM image of the nanowire under study. The location of the GaN/AlN superlattice is outlined with green vertical lines.

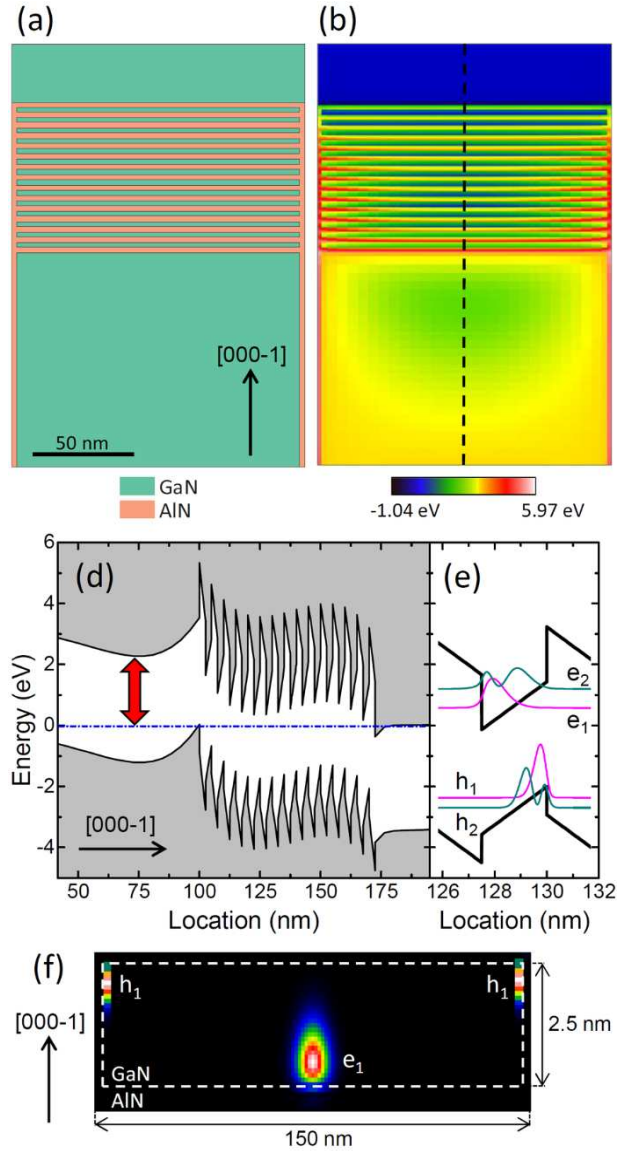

**Figure 5.** Calculation of the band diagram of the nanowire heterostructure in three dimensions. (a) Schematic structure of the wire (left) together with (b) a color-coded cross-section of the conduction band edge viewed along <1–100> (right). (c) Band diagram along the center of the nanowire (dashed line in (b)). The Fermi level is at zero energy (dash-dotted blue line). The red double arrow outlines the shift of the Fermi level with respect to Figure 4(b) due to the presence of surface states in the three-dimensional object. (d) One-dimensional, zero-bias calculation of the band diagram along <0001> of a nanodisk in the center of the stack, with the squared wavefunctions of the first and second electron and hole levels. (e) Three-dimensional, zero-bias calculation of the location of the first electron and hole levels ($e_1$, $h_1$) in a GaN nanodisk. Dashed white lines mark the GaN contour. The disk, located in the center of the stack, is viewed along <1–100>.

# Supplementary Information

The figures in the body of the paper describe the results obtained in one of the nanowires (a typical specimen), where it was possible to perform the complete characterization as a photodetector as well as HAADF-STEM and EBIC. However, the study was validated by the observation of similar results in various nanowires with the same structure. We summarize in this section the results obtained in the various samples, including dark current characteristics (figure S1), photocurrent at zero-bias and 1 V bias (table SI), spectral response under ±1 V bias (figure S2), and HAADF-STEM images (figure S3). Here, NW 4 is the one studied in the main body of the paper.

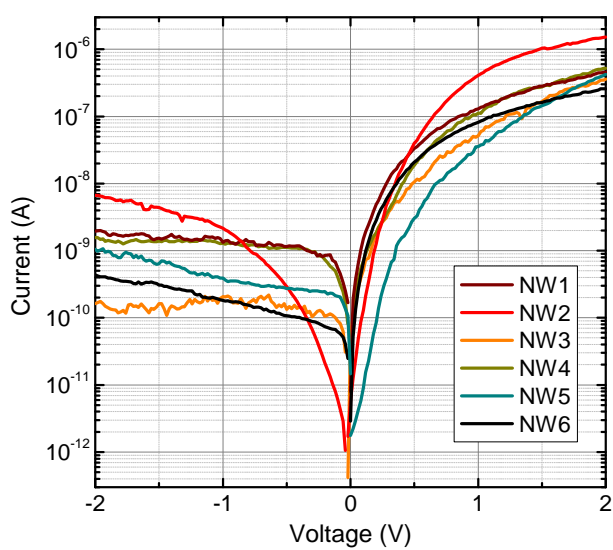

**Figure S1.** Dark current of various nanowires. They systematically display a rectifying behavior with current blocked under negative bias.

**Table S1.** Experimental measurements in the various nanowires under study: dark current ($I_{dark}$) at 1 V bias, and photocurrent ($I_{ph}$) at 1 V bias and at zero bias. Photocurrent measurements were performed under continuous-wave illumination with a HeCd laser (325 nm, power = 1 mW, spot diameter 2 mm). The nanowire whose data is presented in the figures in the body of the paper is NW4 (data in bold).

| Nanowire | $I_{dark}$ (1 V) | $I_{ph}$ (1 V) | $I_{ph}$ (0 V) |
|---|---|---|---|
| NW1 | 130 nA | 4.9 µA | 0.34 nA |
| NW2 | 410 nA | 3.2 µA | 0.48 nA |
| NW3 | 51 nA | 3.4 µA | 3.4 nA |
| **NW4** | **110 nA** | **4.7 µA** | **0.31 nA** |
| NW5 | 36 nA | 2.9 µA | 0.35 nA |
| NW6 | 82 nA | 3.8 µA | 4.44 nA |

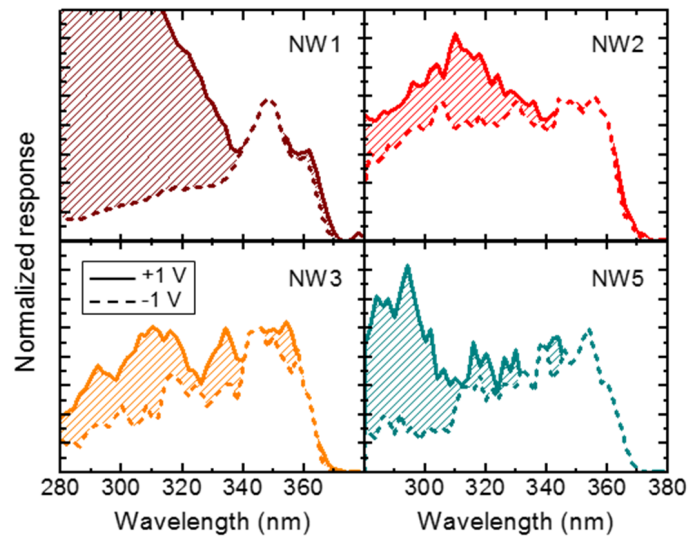

**Figure S2.** Normalized spectral response of various nanowires measured at ±1 V bias.

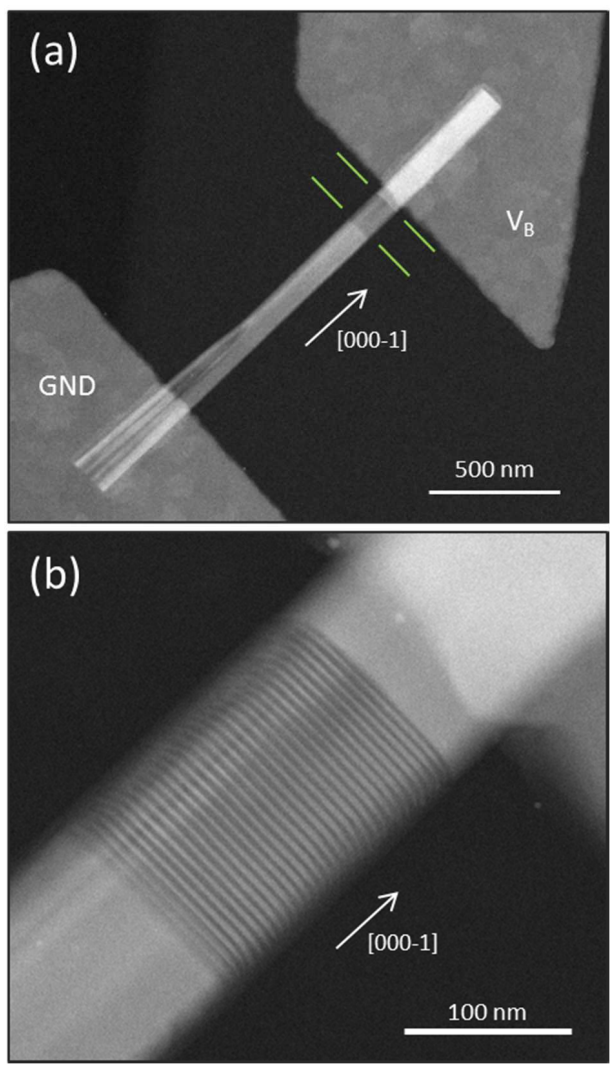

**Figure S3.** HAADF-STEM Images of NW3. The nanowire displays similar features as the nanowire described in Figure 1 (NW4). It results from the coalescence of a bundle of nanowires in the early stages of the growth, well before the deposition of the superlattice. The active region has a diameter of ≈ 130 nm.